\documentclass[aps,prl,twocolumn,superscriptaddress,showpacs]{revtex4}
\usepackage[intlimits]{amsmath}
\usepackage{amsfonts}
\usepackage{graphics}
\usepackage{subfigure}
\usepackage[usenames]{color}
\usepackage{epstopdf,epsfig}
\usepackage{xcolor}
\usepackage{indentfirst}

\usepackage{graphicx}
\usepackage{amsmath}
\usepackage{amsfonts}
\newcommand\be            {\begin{equation}}
\newcommand\EQ           {\begin{equation}}
\newcommand\bea           {\begin{equation}\begin{array}l\displaystyle}
\newcommand\ee            {\end{equation}}
\newcommand\EN           {\end{equation}}
\newcommand\bes           {\begin{subequations}}
\newcommand\esu           {\end{subequations}}

\def\3pt#1#2#3{{\langle{#1}\vert{#2}\vert{#3}\rangle}}

\begin{document}

\title{Prime Suspects in a Quantum Ladder}
\author{Giuseppe Mussardo}
\affiliation{SISSA and INFN, Sezione di Trieste, via Beirut 2/4, I-34151, 
Trieste, Italy}
\author{Andrea Trombettoni}
\affiliation{Department of Physics, University of Trieste,
  Strada Costiera 11, I-34151, Trieste}
\affiliation{SISSA and INFN, Sezione di Trieste, via Beirut 2/4, I-34151, 
Trieste, Italy}
\author{Zhao Zhang}
\affiliation{Tsung-Dao Lee Institute, Shanghai Jiao Tong University, Shanghai 200240, China}
\affiliation{Nordita, KTH Royal Institute of Technology and Stockholm University, Roslagstullsbacken 23, SE-106 91 Stockholm, Sweden}

\begin{abstract} 

\noindent
In this Letter we set up a suggestive number theory interpretation of a quantum ladder system made of ${\mathcal N}$ coupled chains of spin 1/2. Using the hard-core 
boson representation and a leg-Hamiltonian made of a magnetic field and a hopping term, we can associate to the spins $\sigma_a$  the prime numbers $p_a$ so that the chains become quantum registers for square-free integers. The rung Hamiltonian involves permutation terms between next neighborhood chains and a coprime repulsive interaction. The system has various phases; in particular there is one whose ground state is a coherent superposition of the first ${\mathcal N}$ prime numbers. We also discuss the realization of such a model in terms of an open quantum system with a dissipative Lindblad dynamics.   
 \end{abstract}

\maketitle


\noindent
{\bf Introduction}. The aim of this Letter is to point out some interesting connections between quantum many-body systems and number theory, in particular prime numbers. 
Prime numbers are the building blocks of arithmetics and, arguably, one of the pillars of the entire mathematics \cite{Hardy,Apostol}. Their nature has two fascinating but opposite features \cite{Tao}: if their appearance in the sequence of natural numbers is rather unpredictable, their coarse graining properties (e.g. their total number $\pi(x)$ less than $x$) can be captured instead rather efficiently by simple statistical considerations \cite{Kac,Cramer,Erdos-Kac,Schroeder,Julia}. In particular, the scaling 
of the $k$-th prime is particularly plain 
\EQ
p_k \simeq k \, \log k\,\,\,. 
\label{scalingprimes}
\EN
Equally fascinating is the connection between prime numbers and quantum mechanics: Prime numbers, for instance, were the main concern of Shor's algorithm, one of the first quantum
computing algorithms \cite{Shor}. Moreover, the scaling behavior (\ref{scalingprimes}) permits one to show the existence of a single-particle one-dimensional quantum mechanical potential $V(x)$ with eigenvalues given just by the prime numbers and therefore permits one to address the primality test of a natural number in terms of a quantum scattering \cite{GMscattering}: such a potential $V(x)$ can be determined either semiclassically \cite{GMscattering} or exactly, using in this case methods of supersymmetric quantum mechanics \cite{QMprime,SUSYQM}. In experimental setups of cold atom systems, $V(x)$ could be realized using a holographic trap \cite{Cassettari}.

\begin{figure}[t]
\centering
\includegraphics[width=0.17\textwidth]{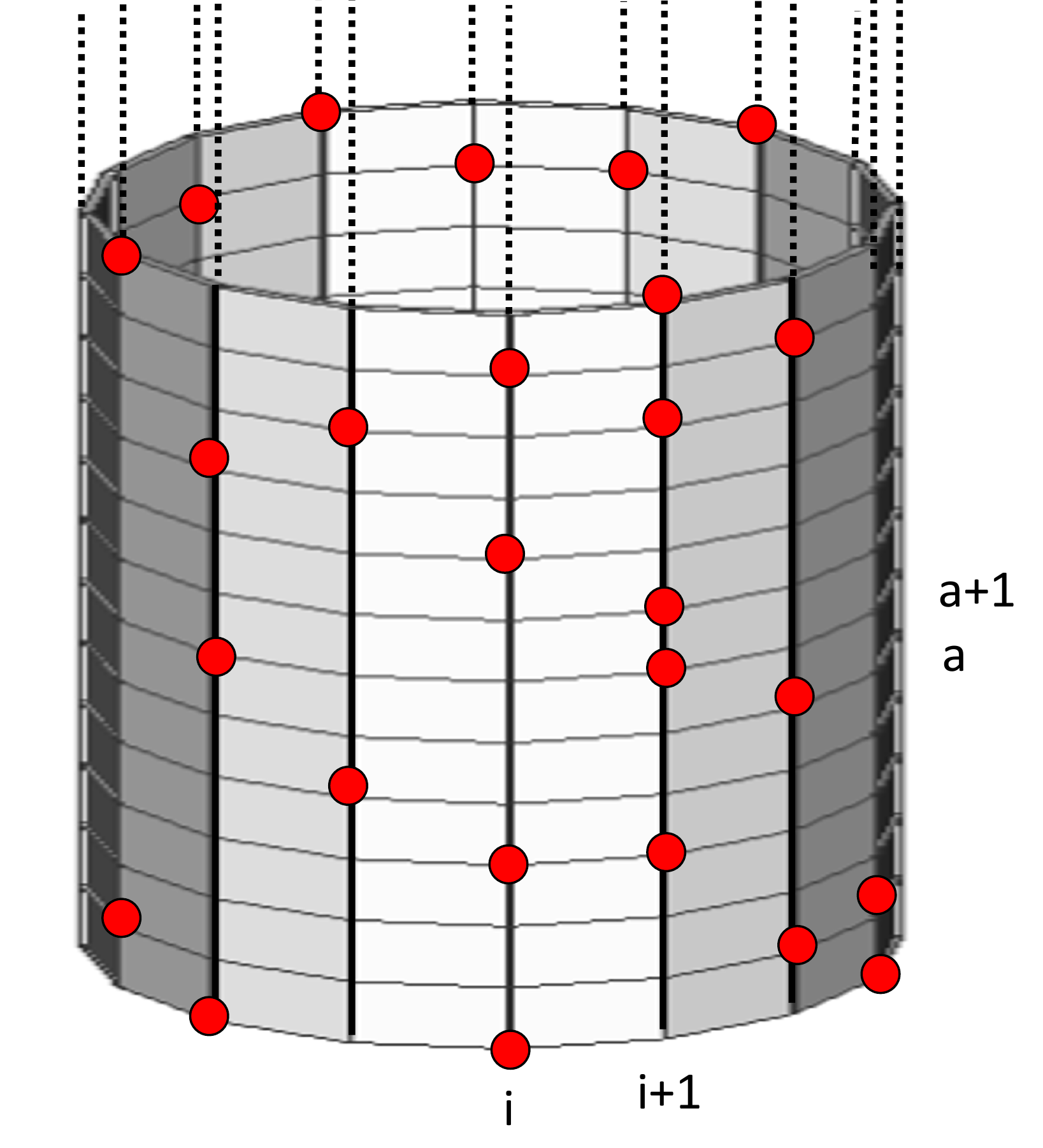}
\caption{Ladder system of ${\mathcal N}$ coupled half-infinite quantum chains of hard-core bosons. Red circles refer to occupied sites.}  
\label{lattice}
\end{figure}

Turning now our attention to quantum many-body systems, for the dense nature of their spectra it is obviously impossible to have energy levels given by prime numbers but we can have instead many-body ground state wave functions expressed in terms of prime numbers. This is what we are going to present below, where we consider a quantum ladder system with a suggestive number theoretic interpretation. We will see that such a system has a rich spectrum of ground states and, in particular, there is one whose wave function is given in terms of a highly coherent superposition of prime number occupations. To the best of our knowledge, this is the first time where a ground state of this type has been constructed.  

Quantum ladder systems, made of coupled one-dimensional chains, have attracted considerable interest in recent years as truly interpolating between one- and two-dimensional systems \cite{Rice,Scalapino,Haldane,Giamarchi,Azuma,Cabra,Ludwig}.  In our case we have ${\mathcal N}$ coupled half-infinite chains of spins 1/2 subjected to a magnetic field 
and a hopping term: as discussed below, properly tuning these two interactions, we can put in correspondence the spins with the prime numbers and reformulate the spin-spin rung interaction in terms of coprimality conditions (two integers are coprime if they do not share common factors other than $1$). 

\noindent
{\bf Degrees of freedom.} As is well known, spin 1/2 can be described by hard-core bosons: the mapping between the Pauli matrices $\sigma_a$ and the 
hard-core annihilation and creation operators $f$ and $f^\dagger$ ($f^2 = (f^\dagger)^2=0$),  is provided by $\sigma_z = f^\dagger f -1/2$; $\sigma_{+}=f^\dagger$; $ \sigma_- = f$ \cite{TDLee}.  Hence, instead of the spins,  we can equivalently take as degrees of freedom the hard-core boson operators $f_i(a)$ and $f^{\dagger}_i(a)$, where the index $i$ refers to the
$i$th chain ($i=1,2,\ldots,{\mathcal N}$), while $a =1,2,\ldots$ to the vertical position along the half-infinite chain (see Fig. \ref{lattice}). Since $(f_i(a))^2 = (f^{\dagger}_i(a))^2 =0$, the occupation number of each vertical site in the ladder can take only values $\{0,1\}$. Let $|{\rm vac} \rangle $ be the vacuum state, i.e.  the state which is annihilated by all the $f_i(a)$'s: for each chain we can then define the state
\EQ
| n_i \rangle \,=\,
\left(\prod_{a=1}^k (f_i^\dagger(a))^{\alpha_a} \right) \, |{\rm vac} \rangle \ \hspace{3mm}, 
\hspace{3mm} 
\,\,\,\alpha_a=\{0,1\}.
\label{local}
\EN
We will show below (see, in particular, Eqs. \ref{PotentialFinale1}-\ref{energyENi}) that it is possible to associate to the $a$th hard-boson along each chain the $a$-th prime number: this allows us to use the notation $|p_a\rangle_i = f_i^\dagger(a) |{\rm vac} \rangle$ and to define for each chain a set of integers whose general form is 
\begin{equation}
n\,=\,p_1^{\alpha_1}\,p_2^{\alpha_2} \cdots \,p_k^{\alpha_k} 
\hspace{3mm}, 
\hspace{3mm} 
\,\,\,\alpha_a=\{0,1\}.
\label{ssff}
\end{equation}
These  are the so-called {\em square-free numbers}, i.e. those integers whose prime factors do not divide them more than once. 
Their first representatives are $n \,= \,2, 3, 5, 6, 7, 10,  11, 13, 14, 15,...$
Remarkably, these numbers are a finite fraction (i.e. $6/\pi^2$)
of all the integers \cite{Schroeder}: indeed, assuming $1/p$ to be the probability that a generic integer is divisible by a prime $p$, the probability that it is not divisible more than once by a prime is given by $ \prod_{p} \left(1-\frac{1}{p^2}\right) \,=\,1/\zeta(2) = \frac{6}{\pi^2}$, where $\zeta(x)$ is the Riemann-zeta function.
Any occupation number configuration of the vertical chains can be associated to a square-free number and viceversa, so each chain plays the role of a quantum register for the square-free numbers. 

It is useful to define 
$
F_i(a) \,=\, f_i^{\dagger}(a)\,f_i(a)\,
$,
(the number operators of the $a$th hard-core boson on the $i$th chain), 
$
F_i \,=\,\sum_{a=1}F_i(a) \,
$
(the total number of hard-core bosons of the $i$-th chain)  and 
$
F^{(a)} \,=\,\sum_{i} F_i(a) \,
$  
(the total number of the $a$ hard-core bosons on the entire ladder lattice). 
It is also convenient to introduce the numbers operators $\hat N_i$ for each chain, such that 
\be
\hat N_i \,| n_j \rangle \,=\, \delta_{i,j} \,n_j \,|n_j \rangle\, \,\,\,.
\label{numberoperators}
\ee
It is worth stressing that the $n_i$'s are {\em not} the true occupation numbers (the actual occupation numbers at each leg are given  by  the $F_i$'s).
The $n_i$'s may be just regarded as useful labels of the hard-core boson degrees of freedom present at each chain: notice that 
there is a one-to-one correspondence between configurations of hard-core bosons and the $n_i$'s since, after all, a decomposition as (\ref{ssff}) not only exists but is also unique. 

The Hamiltonian of our system consists of two terms, relative to rung (R) and leg (L) interactions  
\be
H = \sum_{i=1}^{\mathcal N} \left[H_{i,+1} + H_{i}\right] \,\equiv {\mathcal H}_R + {\mathcal H}_L\,\,\,. 
\label{Hammmm}
\ee

\vspace{1mm}
\noindent
{\bf Rung Hamiltonian}. Let us discuss first the rung Hamiltonian ${\mathcal H}_R$ which acts on a bosonic Hilbert space given by occupation numbers $n_i$'s 
given by the square-free numbers. How to realize such occupation numbers will be discussed later in relation with ${\mathcal H}_L$. Taking for granted 
this Hilbert space, the explicit expression of ${\mathcal H}_R(\lambda)$ is given by ($\lambda \geq 0$)
\be 
{\mathcal H}_R(\lambda) = \sum_{i=1}^{\mathcal N} H_{i,i+1}=\sum_{i=1}^{\mathcal N}  \left[ P_{i,i+1} + \lambda \,  {\mathcal C}_{i,i+1} \right]\,.
\label{hamii+1}
\ee
$P_{i,i+1}$ is the permutation operator between the two n.n. occupation numbers $n_i$ while ${\mathcal C}_{i,i+1}$ is the 
{\em coprimality operator} which counts how many common prime factors are shared between the square-free numbers $n_i$ and $n_{i+1}$. In terms of the 
hard-core boson operators, the coprimality operators can be expressed as ${\mathcal C}_{i,i+1} \,\equiv\,\sum_{a=1}^{\Lambda} F_i(a) F_{i+1}(a)$, where $\Lambda$ is 
a convenient cutoff in the length of the vertical chain (with $\Lambda \rightarrow \infty$ taken first and independently of ${\mathcal N}$). 

${\mathcal H}_R $ 
clearly conserves the total numbers of each $n_i$'s: the Hilbert space is then partitioned in sectors ${\mathcal S}_{\omega_1\ldots \omega_k}(u_1,\ldots,u_k)$, identified by a set of square free numbers $(u_1, u_2,\ldots, u_k )$, with $k \leq {\mathcal N}$ and multiplicities  $(\omega_1,\omega_2,\ldots \omega_k)$ such that $
\sum_{i=1}^{k} \omega_i \,=\, {\mathcal N} $. The dimensions of these sectors are 
$
d(\omega_1,\ldots,\omega_k)\,=\,\frac{{\mathcal N}!}{\omega_1! \,\omega_2! \ldots \omega_k!}
$. 
Even though the number ${\mathcal N}$ of legs may be finite, there are nevertheless infinite sectors, which are obtained by varying both the set of the numbers $u_i$ and their multiplicity $\omega_i$. 

Notice that the symmetry of the system in each of these sectors is $S_{\omega_1} \otimes S_{\omega_2} \cdots \otimes S_{\omega_k}$ rather than $S_N$ ($S_A$ denotes the 
permutation group of $A$ objects). Indeed, the permutation operators $P_{i,i+1}$ of $S_N$ enter directly the Hamiltonian, and, therefore, they do not implement a symmetry of the system, since $P_{i,j}\, {\mathcal H}_R \,P_{i,j}^{-1} \neq H$, where $P_{i,j}$ is a generic operator of $S_N$ which interchanges $n_i$ with $n_j$. So, in light of the actual symmetries of the system, it is natural to consider different $n_i$'s as different species of bosons and require the validity of spin statistics only for particles of the same species (for more details, see Supplemental Material).


\vspace{1mm}
\noindent
{\bf Manifold of the ground states of ${\mathcal H}_R$}. Let us consider the ground states of ${\mathcal H}_R(\lambda)$ by varying the coupling $\lambda$. 

\begin{figure}[t]
\centering
\includegraphics[width=0.50\textwidth]{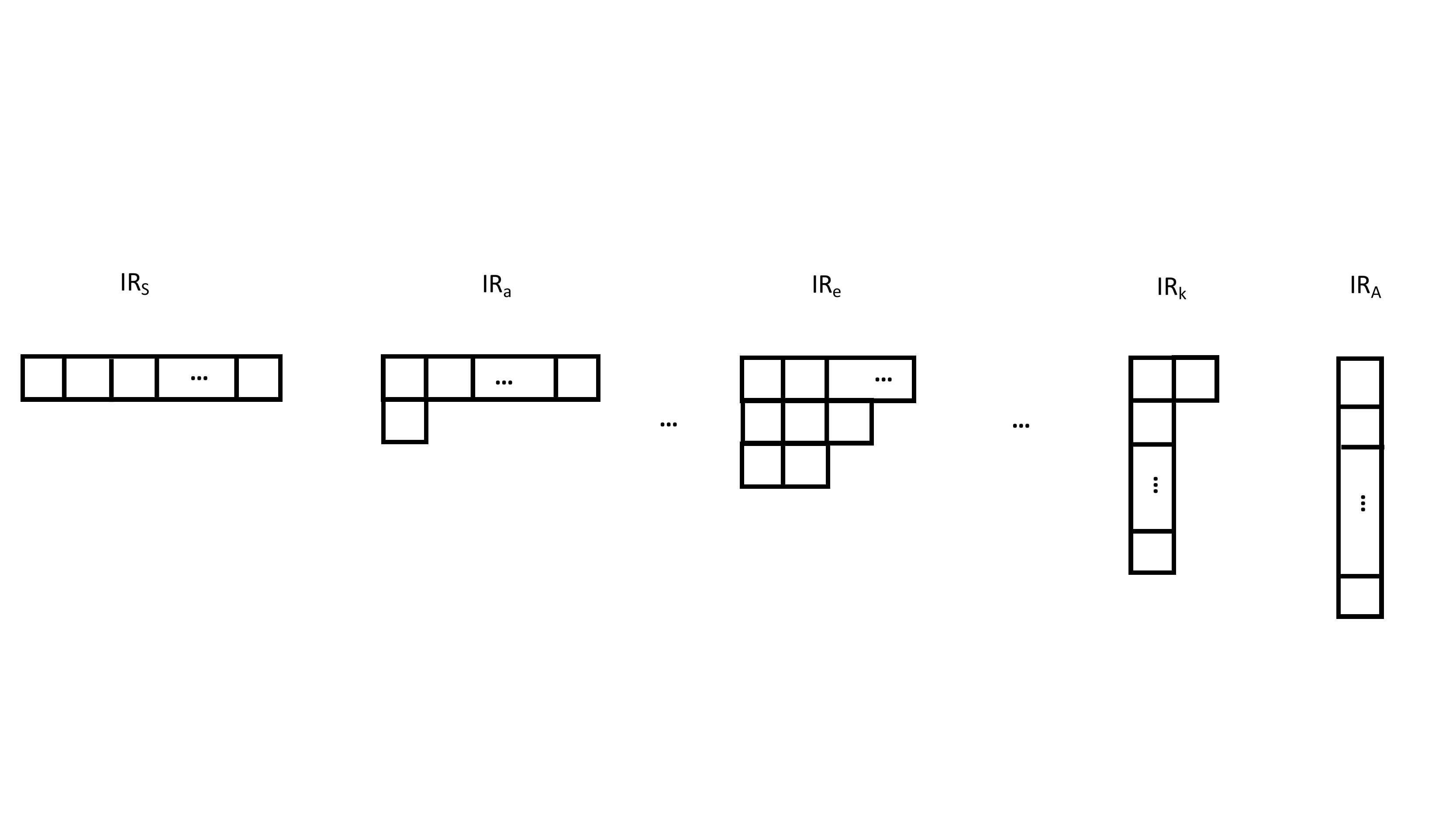}
\caption{Young Tableaux of the symmetrix group ${\mathcal S}_{\mathcal N}$.}
\label{YounTableaux}
\end{figure}

 \textbullet \,\,\,{\em \underline{$\lambda =0$ case}}. When $\lambda =0$, the relative Hamiltonian ${\mathcal H}_R(0)$ can be decomposed in block- orms according to the Irreducible Representation (IR) of the symmetric group ${\mathcal S}_{\mathcal N}$ given by Young Tableaux \cite{YT} and then each block diagonalized separately \cite{footnote2}. While this diagonalization procedure is, in general, highly elaborate, it is instead quite easy to identify the two IRs which give rise to the {\em highest} and {\em lowest} energy states $E =\pm {\mathcal N}$: these are given respectively by the first and the last Young tableaux in Fig.  \ref{YounTableaux}, relative to the fully symmetric and antisymmetric one-dimensional $IR_S$ and $IR_A$. Note that, while $IR_S$ always appears in the decomposition of any sector,  $IR_A$ on the contrary appears only in the decomposition of those sectors where all the $n_i$'s are different numbers. 
The minimum energy $E^*= - {\mathcal N}$ is obtained when each permutation operator $P_{i,i+1}$ can simultaneously take value $-1$: hence, the relative state corresponds to the totally antisymmetric IR (corresponding to the vertically longest Young tableau) and can be written as a Slater determinant built in terms of any set of ${\mathcal N}$ different square-free numbers $|n_a\rangle_i$ 
\EQ
| n^{(1)},\ldots,n^{(\mathcal N)} \rangle_- =\frac{1}{\sqrt{{\mathcal N}!}} 
\left| 
\begin{array}{ccc}
|n^{(1)}\rangle_1 & \cdots &|n^{({\mathcal N})}\rangle_1\\
|n^{(1)}\rangle_2 & \cdots &  |n^{({\mathcal N})}\rangle_2\\
\cdots & \cdots &  \cdots  \\
|n^{(1)} \rangle_{\mathcal N} & \cdots & |n^{({\mathcal N})}\rangle_{\mathcal N}
\end{array}
\right|
\label{Slaterdeterminant}
\EN
One may be surprised that a (hard-)boson wave function is expressed by a Slater determinant but there is nothing strange with this result, since different occupation numbers (as is the case here) correspond to different particle species (i.e. $S_N$ is not a symmetry) and, therefore, there is no violation of spin statistics. Given that we can freely change the square-free numbers involved in 
(\ref{Slaterdeterminant}), the ground state of the Hamiltonian ${\mathcal H}_R(0)$ is then infinitely degenerate, i.e. for any finite number of legs ${\mathcal N}$, the density of these ground states is as much as the density of all states of the Hilbert space. Indeed, let $\Delta$ be a cutoff for the number of square free integers: on a lattice of ${\mathcal N}$ sites, the dimension of the Hilbert space is $D = \Delta^{\mathcal N}$ while the ground states are given by ${\mathcal N}$ different square-free integers, whose number is then ${\tilde d} = \Delta (\Delta -1) (\Delta -2) \cdots (\Delta - {\mathcal N}+1)$. Hence,  $P={\tilde d}/D$ and, taking the limit $\Delta \rightarrow \infty$, we see that $P \rightarrow 1$, independently on the number ${\mathcal N}$ of chains. 
 
 \textbullet \,\,\, {\em \underline{$\lambda >0$ case}}. When we switch on $\lambda >0$, the coprimality term lifts the degeneracy of many of the previous ground states but it leaves several of them untouched: the new set of ground states of ${\mathcal H}_R(\lambda)$ still has $E = - {\mathcal N}$ and corresponds to square-free numbers $n_i$ which have to be not only different but this time also coprime each other! It is indeed the only way to minimize both the permutation and the coprime operators, because in this case the matrix elements of the coprime operator simply vanish. As shown in Supplemental Material, for $\lambda >0$ the fraction of the ground states of the Hamiltonian (\ref{hamii+1}) with respect to the total number of states of the Hilbert space changes radically and is given by  
\EQ
{\rm P}\,=\, 
\prod_{p_a} \left[\left(1 + \frac{{\mathcal N}-1}{p_a +1}\right)\,\left(1 - \frac{1}{p_a +1}\right)^{{\mathcal N}-1}  \right] \,\,\,. 
\label{probabilitygroundstates}
\EN
This quantity now depends on ${\mathcal N}$ and rapidly decreases to $0$ by increasing the number ${\mathcal N}$ of chains, i.e. in presence of the coprimality interaction the number of ground states of ${\mathcal H}_R(\lambda)$  for ${\mathcal N} \gg 1$ is infinitesimally small wrt the dimension of the Hilbert space. 

\vspace{1mm}
\noindent
{\bf Leg Hamiltonian}. Let us now go back to the issue of how one can realise occupation numbers in terms of square-free integers. As mentioned before, the key is to use hard-core bosons but we have to assign them the right energies for putting in correspondence with the prime numbers. This can be achieved by a leg Hamiltonian 
${\mathcal H}_L$ \begin{eqnarray}
&& {\mathcal H}_L = \sum_{i=1}^{\mathcal N} H_i (h,\nu)\,=\,\sum_{i=1}^{\mathcal N} \left[ \sum_{a=1}^{\Lambda} h(a) F_i(a) + \right.\label{PotentialFinale1}\\
&&\left. \hspace{2mm} -  \frac{1}{2} \,\sum_{a \neq b} J_{ab} \, \left(f^\dagger_i(a) f_i(b) + f^\dagger_i(b) f_i(a) \right)\right]
\,\,\,, \nonumber
\end{eqnarray}
which simply involves a magnetic field $h(a)$ and a hopping term $J_{ab}$  properly tuned 
\EQ
 h(a) = h\, \log\, p_a\,
\hspace{3mm}
,
\hspace{3mm}
J_{ab} = J_{|a-b|} =\nu/|a-b|\,\,\,,  
\label{logpa}
\EN
($h, \nu >0$ are two coupling constants). As shown in detail in Supplemental Material, the increasing values of the magnetic field $h(a)$ along the chain according to the logarithm of the primes dictate the long-range nature of the hopping term $J_{ab}$. It is precisely thanks to this long-range dependence of $J_{ab}$ that the ratio $h/\nu$ of the two couplings truly captures the competition present in ${\mathcal H}_i$: indeed, when $h/\nu \rightarrow \infty$, the eigenstates of  ${\mathcal H}_i$ are given by the $|p_a\rangle_i$'s (in this case they are ``localised'' on the primes). The magnetic contribution of a state as (\ref{local}) in the Leg Hamiltonian is given by 
\EQ
M_{n_i} \,=h\, \, \sum_{a=1}^k \alpha_a \, \log (p_a)_i \,=\,h\, \log n_i \,\,\,.
\label{energyENi}
\EN
and these values are {\em never} degenerate for the unique decomposition in terms of primes of any number $n_i$ (see eq.\, (\ref{ssff})). When $h/\nu \rightarrow 0$, the eigenstates are instead ``delocalised" along the whole chain: with periodic b.c. ($J_{ab}$ is a circulant matrix in this case), the ground state of $H_i$ is the so-called Prime State \cite{Latorre-Sierra} 
\EQ
|{\mathbb P}_0\rangle_{\Lambda} \,=\,
\frac{1}{{\sqrt{\Lambda}}}\, \left( 
|p_1 \rangle + |p_2 \rangle + \cdots |p_{\Lambda}\rangle \right) \,\,\,,
\label{primestate}
\EN 
which is completely delocalized in the space of the primes, as are also delocalized the excited states ($\omega = e^{2 \pi i/\Lambda}$)
\EQ
|{\mathbb P}_k\rangle \,=\, 
 \frac{1}{{\sqrt{\Lambda}}}\,
  \left( 
|p_1 \rangle +\omega^{1 k}\, |p_2 \rangle + \cdots \omega^{(\Lambda -1) k}  \,
|p_{\Lambda}\rangle \right) \,\,,. .
\label{exciteddelocalised}
\EN

\vspace{1mm}
\noindent
    {\bf Phases of the system}. The phase diagram of the full Hamiltonian (\ref{Hammmm}) is quite rich. Let us briefly discuss some of its cases. 
 It is easy to see that, taking $\nu \rightarrow \infty$ (keeping all other coupling constants fixed), the system goes into a ``stripe phase'' described by the factorized ground state made of the
    Prime States (\ref{primestate})
\be
|\Psi\rangle_0 \,\simeq\, \otimes_i \,|{\mathbb P}\rangle_i\,\,\,. 
\label{stripes}
\ee
Expanding each $|{\mathbb P}\rangle_i$ in the prime basis, this ground state is made of equally weighted vectors of all possible sectors of the theory, whose 
degeneracy will be eventually solved by taking into account the coprimality interaction and the magnetic field ($\lambda$ and $h$ both small compared to $\nu$). In this phase, factorized expressions also hold for the excited states, $|\Psi\rangle_k \simeq \otimes_i |{\mathbb P}_k\rangle_i$. Assuming periodic boundary conditions and a cutoff $\Lambda$ along each chain, for the ground and excited state energies of this phase we get  $E_k \simeq -\nu\, {\mathcal N} e_k$ where, for $\Lambda \rightarrow \infty$
\begin{eqnarray}
e_0&\simeq & - 2 \, \left(\log \frac{\Lambda}{2} + \gamma_{E} + \frac{1}{\Lambda} + \cdots \right) \nonumber\,, \\
e_k &\simeq  & \log\left[1- \cos \left(\frac{2 \pi k}{\Lambda}\right)\right] + \log 2\nonumber \,,
\end{eqnarray}
which can be made finite by subtracting the leading divergent term $\log \Lambda/2$. 

Taking instead $h \rightarrow \infty$ (and neglecting for simplicity the hopping term in ${\mathcal H}_L$), the system goes into its ``ordered phase'', characterized by an occupation number at each chain given by the lowest prime $p_1 =2$ 
\be
| \Psi \rangle_0 \,\simeq\, \otimes_{i=1}^\mathcal{N}\, | 2 \rangle_i \,\,
\label{ordered}
\ee
and ground state energy $E_0^{({\rm ord})} \simeq {\mathcal N} (h \log 2 + 1 + \lambda)$. For small $h$, the ground state (\ref{ordered}) is, however, unstable with respect to  the proliferation of other numbers $n_i$ (which may replace some of the $2$'s present). Indeed, when such numbers $n_i \neq 2$ exist in some of the chains, the ground state energy tends to decrease for: (i) the presence of other $IR$'s in the decomposition of the permutation term of the Hamiltonian (in addition to $IR_S$, the only IR present in the ordered phase); typically these $IR$'s have lower energy than $E = {\mathcal N}$ (the rule of thumb being, the longer the Young tableau in the vertical direction, the lower the corresponding minimum energy in that IR); (ii) a lower contribution coming from the coprimality term, since there are less pairs of equal particles. Imagine, for instance, replacing one of the $2$'s in the ordered phase with a $3$: the new IR needed in this case is the second Young tableau (from left) in Figure \ref{YounTableaux} which, with periodic boundary conditions, has dimension ${\mathcal N}$ and spanned by the ${\mathcal N}$ vectors ($m=1,2,\ldots,{\mathcal N}$) 
\be
| m \rangle \,\equiv \,| 2, 2, 2, \ldots,  \underset{\underset{m'th\ chain}{\uparrow}}{{3}}, \ldots, 2, 2, \ldots\rangle\,.
\label{basis}
\ee 
The number $3$ plays a role of a defect w.r.t. the ordered ground state. On the space spanned by these ${\mathcal N}$ vectors, the term $\sum_{i}^{\mathcal N} P_{i,i+1}$ in the Hamiltonian has 
$|v_k\rangle = 1/\sqrt{\mathcal N} e^{i k m} |m\rangle$ as eigenvectors and spectrum given by $\hat E_k = {\mathcal N} - 2 + 2 \cos 2\pi k/{\mathcal N}$, whose minimum is $\hat E_{min} = ({\mathcal N} -4)$. The expectation value of the coprimality operators on the $|v_k\rangle$ eigenvectors is simply (${\mathcal N}-2) \lambda$. So,  putting together the two terms, the minimum energy in the defect sector is $E_{min}^{({\rm def})} = {\mathcal N} - 4 + \lambda ({\mathcal N} - 2) + {\mathcal N} \,h \, \log 2 + h \,\log (3/2)$ 
(neglecting for simplicity the hopping term in ${\mathcal H}_L$). Comparing now $E_0^{({\rm ord})}$ with $E_{min}^{({\rm def})}$, we can determine the minimum value of $h$, i.e. $h_c$, for which the ordered phase is indeed stable 
\be
E_0^{({\rm ord})} \leq E_{min}^{({\rm def})} 
\,\hspace{5mm}
{\rm if} 
\hspace{5mm} h_c \, \log (3/2) \geq 4 + 2 \lambda 
\ee
Let us finally consider the limit in which $\lambda \rightarrow \infty$ and $h \rightarrow 0$ (in a way determined below), also imposing the extra condition $\nu \ll h$. In this case, the system goes into a ``prime-number phase'', which consists in minimizing simultaneously the permutation and the coprimality operators, 
adjusting accordingly the magnetization operators. A state which satisfies all these requirements consists of the Slater determinant of the first ${\mathcal N}$ primes. As for a fermionic system, also for hard-core bosons this condition defines a ``Fermi energy" given by filling the first ${\mathcal N}$ levels and its value is 
\EQ
 E_F \,=\, \sum_{a=1}^{\mathcal N} \log p_a \,=\,\log\, \left(\prod_{a=1}^{\mathcal N} p_a \right) \,=\,\log\, \tilde P({\mathcal N}) \,\,\,,
 \EN 
where $\tilde P({\mathcal N})$ is the {\em primorial}, i.e. the product of the first  ${\mathcal N}$ consecutive prime numbers. Since this quantity goes asymptotically as $ \tilde P({\mathcal N}) \simeq e^{p_{\mathcal N}} \simeq e^{{\mathcal N} \log {\mathcal N}}$ \cite{ruiz}, we have 
$E_F \,\simeq {\mathcal N} \,\log {\mathcal N} 
 \label{Fermienergyasymp}
$. 
Hence the ground state energy in the ``prime phase'' is given by 
\be
E_0^{(prime)} \,=\, -{\mathcal N} + h\, {\mathcal N} \, \log {\mathcal N} \,\,\,. 
\label{e0prime}
\ee
So, letting $h$ vanish as $h \simeq \tilde h/\log{\mathcal N}$ for ${\mathcal N}\rightarrow \infty$ we have a ground state energy of the ``prime phase'' which scales linearly with 
the number ${\mathcal N}$ of chains.  Hence, this state gives rise to a ``Fermi surface" in terms of the first ${\mathcal N}$ prime numbers. These primes are simultaneously present on each 
chain, being spread on the entire ladder system, although quantum coherently assembled by a Slater determinant, see eq.\,(\ref{Slaterdeterminant}).  As for other Fermi surfaces, there are soft modes above this ground state it: indeed, if we replace one prime number $p_c= p_{{\mathcal N} - \epsilon}$ (inside and close to the Fermi surface) with other one $p_e = p_{{\mathcal N} + \delta}$ (placed outside and close to it), the variation of the energy of the corresponding wave functions is simply 
\be
\Delta E \,=\,\tilde h \, \log \frac{p_e}{p_c} \simeq \tilde h\, \frac{\delta + \epsilon}{\mathcal N} \,,
\,\,\,
\label{gap}
\ee 
where we have used the scaling law (\ref{scalingprimes}). So, for a finite ${\mathcal N}$, the system has a gap which, however, scales to zero as
$1/{\mathcal N}$ if we send the number of chains to infinity. 

\vspace{1mm}
\noindent
{\bf Lindbladian dynamics}. 
 A natural question is how the system is able to reach one of its ground states, say the ``prime ground state'' given by the coherent superposition of the first ${\mathcal N}$ primes. 
 One way is to set up a dissipative dynamics able to efficiently ``filter'' such a ground state starting from an initial configuration made of an arbitrary mixture of excited states. 
 This procedure can be implemented by choosing a suitable and optimized set of Lindblad operators (see, e.g. \cite{Petruccione}) which induce a dissipative dynamics 
 for the density matrix $\rho$ of our system ruled by the master equation 
\begin{equation}
\dot{\rho}\,=\, -\,\frac{i}{\hbar} [H,\rho] + {\cal L}[\rho]\,\,\,,
\label{eq:rho}
\end{equation}
where $H$ is the ladder Hamiltonian (\ref{Hammmm}) while ${\cal L}[\rho]$ is the Lindbladian term describing spontaneous emission processes. Of course one has to specify the Lindbladian operator, a goal achieved by posing 
$${\cal L}[\rho]=\sum_{i=1}^{{\cal N}} \left( \gamma_i L_i \rho L_i^\dag - (\gamma_i/2) \{L_i^\dag L_i, \rho \}\right)\,\,\,, $$ and identifying a suitable set 
of the $L_i$'s operators.

For our purposes, notice that the quantum superposition is induced by the rung Hamiltonian ${\cal H}_R$ while the intra-leg term ${\cal H}_L$ permits the hopping, i.e. 
the reshuffling, of the particles among the $|p_a\rangle$ levels inside each of the chains. Hence, if our aim is to target the ground state made of the first ${\mathcal N}$ primes, it is sufficient   
to choose for $L_i$ the following operators 
\begin{equation}
L_i \,=\, \sum_{a>{\cal N} } \sum_{b \le {\cal N}} f_i(a) f_i(b)^\dag \,\,\,, 
\label{form}
\end{equation}
(in principle, one could also consider level-dependent coefficients $\gamma_i^{(ab)}$). The rationale behind this choice 
is that the dissipative term does not act in the space of the first ${\mathcal N}$ levels while at the same time does favor the occupation of such a subspace. We expect an interplay between the term $\propto \nu$ present in ${\mathcal H}_L$ and the Lindbladian term ${\mathcal L}$, in the sense that a nonvanishing $\nu$ tends to decrease the characteristic time in which the system reaches our target subspace.

\noindent
{\bf Conclusions.} Number Theory is the paradigmatic example of pure mathematics. Yet the theory of integers can appear totally unexpected in quantum mechanics systems, 
providing new perspectives on their dynamics. In this paper we have considered a many-body quantum ladder system, made of ${\mathcal N}$ coupled quantum chains, whose degrees of freedom and interactions have a very direct interpretation in terms of prime numbers and basic properties thereof. We have shown that such a system has many different phases. Among the major capabilities of this system there is the possibility of realizing a ground state made of a coherent superposition of the first ${\mathcal N}$ primes.

\vspace{1.1mm}
 
\begin{center}
{\bf Acknowledgments}
\end{center}

\noindent
We thank Donatella Cassettari, Haiyuan Zou, Lorenzo Piroli and Jacopo Viti for interesting discussions. GM acknowledges the grant PRIN 2017E44HRF.

\newpage

\clearpage 
\setcounter{equation}{0}
\renewcommand{\theequation}{S\arabic{equation}}
\setcounter{page}{1}
\setcounter{figure}{0}
\onecolumngrid

\begin{center}
{\Large{\bf Supplementary Material}}
\end{center}

\section{The Hamiltonian based on Permutations}
In this Section we enlighten some properties of the Hamiltonian ${\mathcal H}_R$ given in eq.\,(6) of the text to better understand the origin of the 
Slater determinants given in eq. (7) in the text as its ground state wave-functions. To this aim it is actually sufficient to focus the attention on the Hamiltonian obtained when the coupling constant $\lambda$ of the coprimality operator $C_{i,i+1}$ vanishes, namely 
\begin{equation}
H = \sum_{i=1}^{\mathcal N}  P_{i,i+1}\,\,\,,
\label{Suther}
\end{equation}
where $P_{i,i+1}$ is the permutation operator between the two n.n. ``occupation numbers" $n_i$. Let's notice that 
\begin{itemize}
\item
the permutation operators (which usually describe a symmetry of a quantum Hamiltonian under permutations) enter instead the Hamiltonian itself! 
\item therefore the permutation group $S_N$ is \underline{{\em not}} a symmetry of the Hamiltonian, i.e. $P_{i,j} H P_{i,j}^{-1} \neq H$, where $P_{i,j}$
is a generic permutation operator.  The true symmetries of the Hamiltonian (1) are discussed below. 
\end{itemize}
As suggested by Sutherland [27], the different occupation numbers on which it is based the Hilbert space have to be considered as different species of particles. With this observation in mind, 
let's show that, in a one-dimensional lattice where there are bosonic degrees of freedom,  the ground state(s) of the Hamiltonian (\ref{Suther})  
are given indeed in terms of Slater determinants, as in eq. (7) of the main text. It is important to notice that the Hamiltonian has a $+$ sign in front and therefore tends to select totally anti-symmetric states as its ground states.  
\begin{enumerate}
\item ${\mathcal H}_R $ clearly conserves the total numbers of
  each $n_i$'s: this implies that the Hilbert space splits in different
  sectors ${\mathcal S}_{\omega_1\ldots \omega_k}(u_1,\ldots,u_k)$, identified by a set of 
  numbers $(u_1, u_2,\ldots, u_k )$, with $k \leq {\mathcal N}$ and multiplicities  $(\omega_1,\omega_2,\ldots \omega_k)$ such that $
  \sum_{i=1}^{k} \omega_i \,=\, {\mathcal N} $. In the case of interest for our paper the numbers $u_i$'s are square free numbers.
  
Each sector has then its representative state, which is given by writing the occupation numbers $u_i$'s from lower to higher values, as 
\begin{equation}
|\underbrace{u_1, \ldots, u_1}_{\omega_1}, \underbrace{u_2,\ldots, u_2}_{\omega_2}, \ldots, \underbrace{u_k,\ldots,u_k}_{\omega_k}\,\,\rangle \,\,\,,
\label{representative}
\end{equation}
with $u_1 < u_2 < \ldots u_k$. Using the representative state, the remaining states of a given sector are obtained by permuting in all possible inequivalent ways its numbers. The dimensions of these sectors are 
\begin{equation}
d(\omega_1,\ldots,\omega_k)\,=\,\frac{{\mathcal N}!}{\omega_1! \,\omega_2! \ldots \omega_k!}\,\,\,.
\end{equation}
\item Restricted to a given sector ${\mathcal S}_{\omega_1\ldots \omega_k}(u_1,\ldots,u_k)$, the Hamiltonian is invariant under the group
\begin{equation}
S_{\omega_1} \otimes S_{\omega_2} \cdots S_{\omega_k}
\end{equation}
where $S_A$ denotes the permutation group of $A$ objects. This is actually the only 
symmetry that has to be respected by the eigenvectors of the Hamiltonian in each sectors. As we commented previously, the permutation group $S_N$ is, on the contrary,  
not a symmetry of the system: as a matter of fact, the permutation operators were actually used to generate the Hilbert space of a given sector, acting on the reference state. 
\item In any given sector 
the permutation operators $P_{i,i+1}$ has a matrix representation of dimension $d$ equal to the dimension of the sector, $d= d(\omega_1,\ldots,\omega_k)$. 
However, such a representation is in general {\bf reducible}. Decomposing the representation of dimension $d$ of the permutation operators in terms of the Irreducible Representations associated to the Young Tableaus, the Hamiltonian takes the block form shown in Figure \ref{blf2}
\begin{figure}[h!]
\centering
\includegraphics[width=0.2\textwidth]{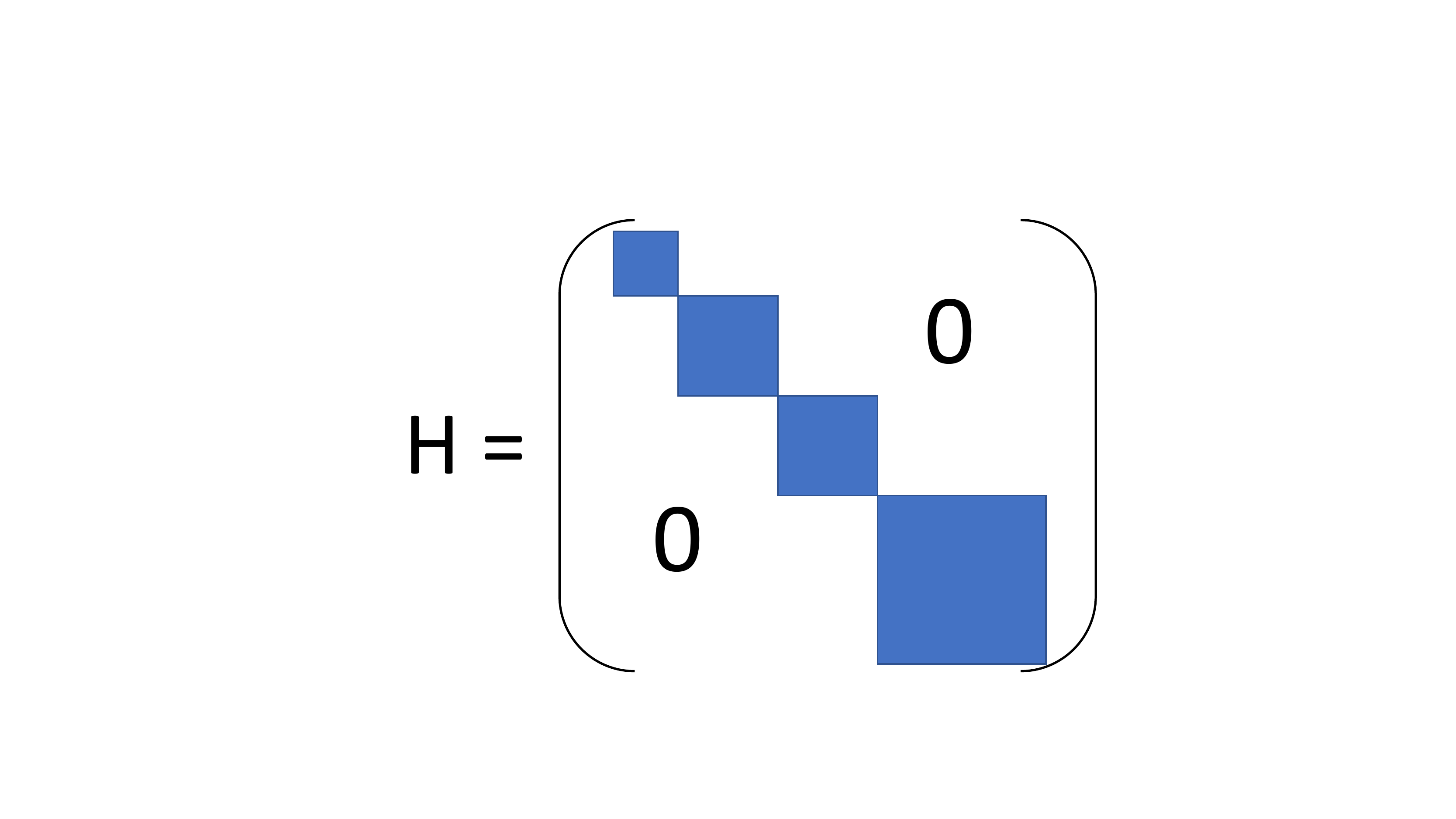}
\caption{Block form of the Hamiltonian (\ref{Suther})}
\label{blf2}
\end{figure}
\item The dimension of each block is equal to the dimension of the Irreducible
  Representations
  of the Young Tableaus and each block can be diagonalised separately.
\item So, the question to find the minimum eigenvalues (and the
  corresponding eigenvector) of the Hamiltonian boils down to find out
  which is the Irreducible Representation 
(i.e. the block) with the minimum eigenvalue. 
\item Since each permutation operator $P_{i,i+1}$ {\em individually}
  has eigenvalues $\pm 1$, the maximum of the Hamiltonian is
  $E_{max} = ({\mathcal N}-1)$ and the minimum  $E_{min}=- ({\mathcal N}-1)$
  (with open b.c.). Clearly these two values can only be obtained if
  the Irreducible Representations are $1$-dimensional: the only 1-d
  Irreducible Representations are the totally symmetric one
  and total anti-symmetric one. 
\item The totally symmetric representation exists in the IR decomposition of any sector. Indeed
  it is sufficient to construct the totally symmetric linear combination
  of the vectors generated by the representative state (\ref{representative}). This is actually  the {\em only} state which is symmetric under $S_d$. 
\item However, the totally anti-symmetric Irreducible Representation
  only exists for those sectors where all occupation numbers are different.  Being all occupation numbers different (and considering them as different species), there is obviously no problem
  with the spin-statistic property of the corresponding wave-function. 
\item If some occupation numbers were instead equal, the totally anti-symmetric Irreducible Representation simply does not exist in the IR decomposition of the corresponding sector. 
\end{enumerate}

\vspace{5mm}
\noindent
{\bf Two explicit examples}. Two simple examples make explicitly clear what discussed above


\vspace{3mm}
\noindent 
{\em \underline{First example}}. As a first example, let's consider a chain of only $3$ sites (with open boundary conditions). The Hamiltonian is then in this case  
\begin{equation}
H = P_{12} + P_{23} 
\label{ham123}
\end{equation}
Let's now consider the sector in which all the occupation numbers at the three sites are different, say $2, 3, 5$. The Hilbert space has dimension $d=6$ and spanned by the vectors 
\begin{eqnarray}
&& |v_1\rangle = |2 \rangle_1 |3\rangle_2 | 5\rangle_3
\hspace{3mm}
,
\hspace{3mm}
|v_2\rangle = |2\rangle_1 |5\rangle_2 | 3\rangle_3
\hspace{3mm}
,
\hspace{3mm}
|v_3\rangle = |3 \rangle_1 |2 \rangle_2 | 5\rangle_3\\
&& 
|v_4\rangle = |5 \rangle_1 |3\rangle_2 | 2\rangle_3
\hspace{3mm}
,
\hspace{3mm}
|v_5\rangle = |5\rangle_1 |2\rangle_2 | 3\rangle_3
\hspace{3mm}
,
\hspace{3mm}
|v_6\rangle = |3 \rangle_1 |2 \rangle_2| 5\rangle_3
\end{eqnarray}
Notice that in this sector there is no permutation symmetry at all (the group $S_1 \otimes S_1 \otimes S_1$ is of course trivial). 
 The decomposition of this $6$-dimensional representation (which is nothing else but the adjoint representation) is in terms of the Young Tableaux shown below. \begin{figure}[h!]
\centering
\includegraphics[width=0.5\textwidth]{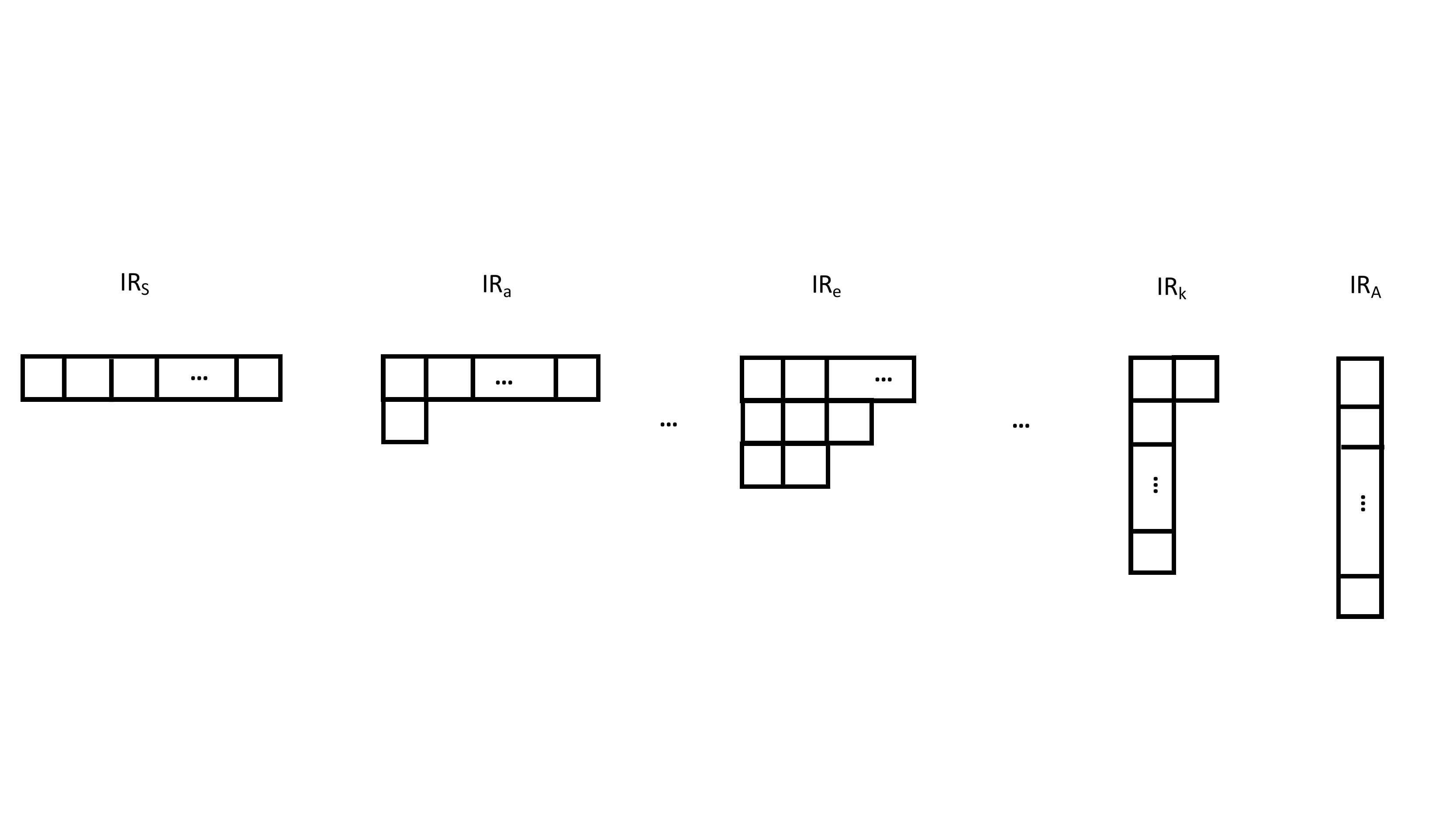}
\label{blockform2}
\end{figure}

The first and the last Irreducible Representation enter one time in the block decomposition, while the middle Irreducible Representation enters twice and therefore we have the sum rules valid for the Irreducible Representations entering the adjoint representation 
\begin{equation}
d_S^2 + d_b^2 + d_A^2 \,=\, 6
\end{equation}

\noindent
The ones which we are mostly concerning here are the first and the last, which are both one-dimensional Irreducible Representations. 
The first Young Tableau (on the left) corresponds to the totally symmetric combination 
\begin{equation}
|S\rangle \,=\, \frac{1}{\sqrt{3!}} \left(|v_1\rangle + |v_2\rangle + |v_3\rangle + |v_4\rangle + |v_5\rangle + |v_6\rangle\right)
\end{equation}
which is eigenvector of the Hamiltonian (\ref{ham123}) with the maximum eigenvalue $E_{max} = 2$, while the last Young Tableau (on the right) corresponds to the combination 
\begin{equation}
|A\rangle \,=\, \frac{1}{\sqrt{3!}} \left(|v_1\rangle - |v_2\rangle + |v_3\rangle - |v_4\rangle + |v_5\rangle - |v_6\rangle\right)
\end{equation}
which is eigenvector of the Hamiltonian (2) with the minimum eigenvalue $E_{min} = - 2$. Notice that $|A \rangle$ can be written as a Slater determinant
\begin{equation}
| A \rangle =\frac{1}{\sqrt{3!}} 
\left| 
\begin{array}{ccc}
|2\rangle_1 & |2 \rangle_2 &|2 \rangle_3\\
|3\rangle_1 & |3 \rangle_2 &|3 \rangle_3\\
|5\rangle_1 & |5 \rangle_2 &|5 \rangle_3
\end{array}
\right|
\label{Slaterdeterminant}
\end{equation}
The state $|A \rangle$ is the only one which minimises {\em both} $P_{12}$ and $P_{23}$ entering the Hamiltonian. 
\vspace{3mm}

\vspace{3mm}
\noindent 
{\em \underline{Second example}}. As a second example let's now consider once again a chain of only $3$ sites (with open boundary conditions) with the same Hamiltonian (\ref{ham123}), but in the sector where there are two {\em equal} occupation numbers, say $2,2,3$. The Hilbert space in this case has dimension $d=3$ and spanned by the vectors 
\begin{equation}
 |w_1\rangle = |2 \rangle_1 |2\rangle_2 | 3\rangle_3
\hspace{3mm}
,
\hspace{3mm}
|w_2\rangle = |2\rangle_1 |3\rangle_2 | 2\rangle_3
\hspace{3mm}
,
\hspace{3mm}
|w_3\rangle = |3 \rangle_1 |2 \rangle_2 | 2\rangle_3
\label{basis}
\end{equation}
These vectors are clearly symmetric under those permutation operators which exchange the two equal occupation numbers but not for the other permutation operators. 
In other words, the symmetry group is in this case $S_2 \otimes S_1$. 
The decomposition of this $3$-dimensional representation is in terms of the
two Irreducible Representations shown below of dimensions $d_S=1$ and $d_b=2$. 
\begin{figure}[h!]
\centering
\includegraphics[width=0.25\textwidth]{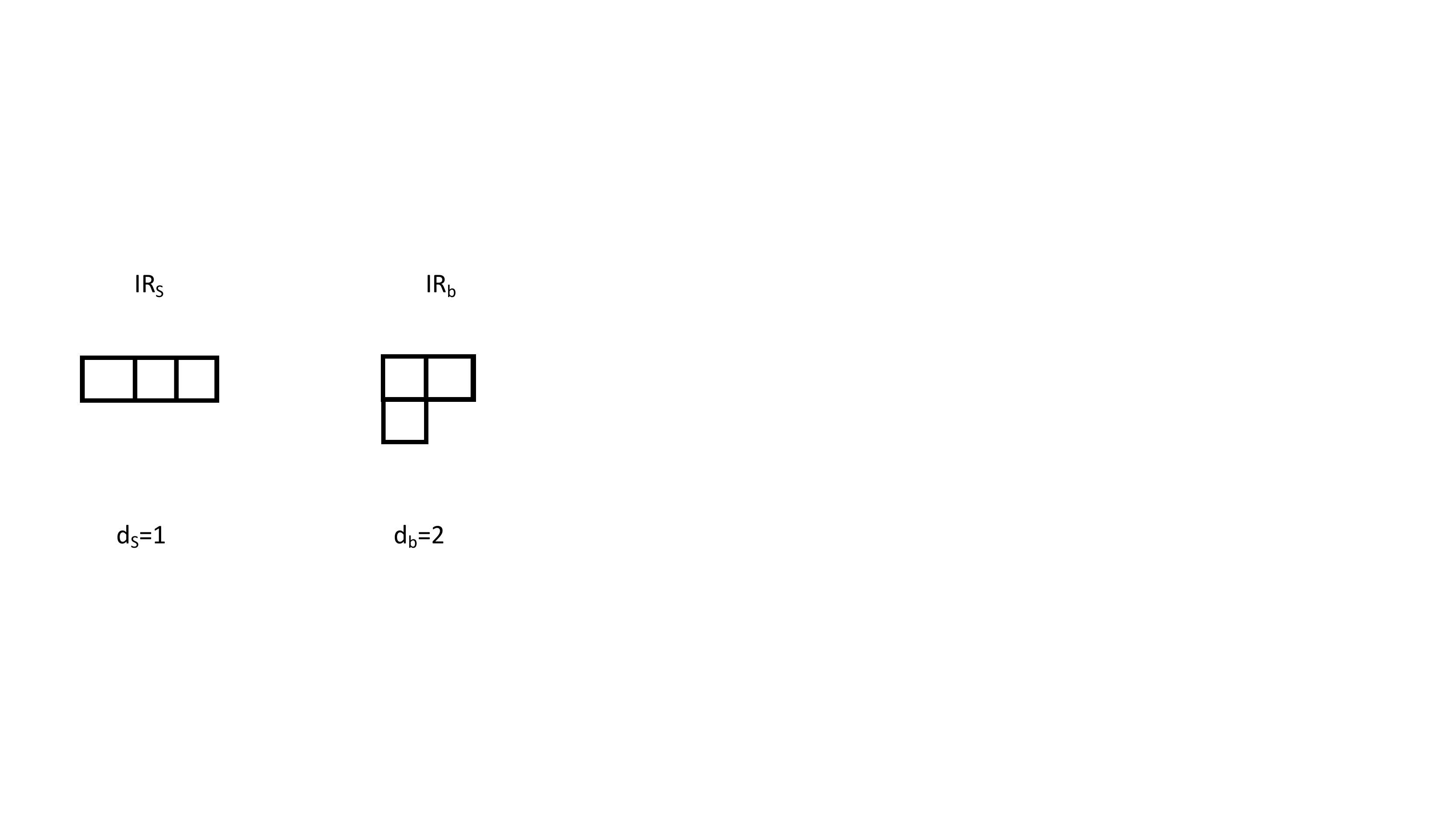}
\label{blockform3}
\end{figure}

\noindent
The Hamiltonian in the basis (\ref{basis}) is given by the $3\times 3$ matrix 
\begin{equation}
H\,=\,
\left(\begin{array}{ccc}
1 & 1 & 0 \\
1 & 0 & 1 \\
0 & 1 & 1
\end{array}
\right)\,\,\,.
\end{equation}
In the new basis, given by the vectors identified by the Young Tableaux, the Hamiltonian becomes of block form: one block has dimension $d=1$ while the 
another one has dimension $d=2$. 

The eigenvalue of the one-dimensional block is $E_{max} =2$ and the corresponding eigenvector is the totally symmetric combination 
\begin{equation}
|{\mathcal S}\rangle \,=\,\frac{1}{\sqrt{3}} \left(|w_1 \rangle + |w_2\rangle + w_3\rangle\right)
\end{equation}
The $2\times 2$ matrix has instead 2 different eigenvalues and eigenvectors 
\begin{eqnarray}
&& E_1 = 1 \hspace{6mm } \rightarrow \hspace{3mm} |E_1\rangle \,=\,\frac{1}{\sqrt{2}} (|w_1 \rangle - |w_3\rangle) \\
&& E_2 = -1 \hspace{3mm } \rightarrow \hspace{3mm} |E_2\rangle \,=\,\frac{1}{\sqrt{6}} (|w_1 \rangle - 2 |w_2\rangle + |w_3 \rangle) \nonumber
\end{eqnarray}  
Notice that in the IR decomposition of the sector ${\mathcal S}_{2,1}(2,3)$ it is absent the totally anti-symmetric representation (corresponding to a vertical Young diagram with 3 boxes). Morever, except the totally symmetric eigenvector $ |{\mathcal S}\rangle$, the other two eigenvectors do not have, correctly, any $S_3$ symmetry but they only have the $S_2 \otimes S_1$ symmetry which interchange the sites where there are equal occupation numbers. 

\section{ Probability of divisibility by $p_a$ of a square-free number}

\begin{table}[b]
\begin{center}
\begin{tabular}{|l|l|l|}\hline
  primes  &  
  numerical & theoretical \\
$p_a$ & probability & probability  \\\hline
$2 $ & $0.333331$ & $0.333333$  \\
$3 $ & $0.249998$ & $0.250000$\\
$5$ & $0.166670 $ & $0.166666$\\
$7 $ & $0.1250001 $ & 0.125000\\
$11$ & $0.0833331$ & $0.0833333$ \\
$13 $ & $0.0714281$ & $ 0.0714286$\\
$17$ & $0.0555547$ &$0.0555556$ \\ 
\hline
\end{tabular}
\end{center}
\label{tablepppp}
\caption{
  Numerical vs theoretical probability of divisibility of square-free numbers by a prime. The 
  numerical data are given by the ratio $N_a/N$, where $N=10^7$ is the number of square-free numbers considered and $N_a$ is the number of them divisible by the prime $p_a$. The theoretical probability is $1/(p_a +1)$. }
\end{table}

In order to determine the probability (12) given in the text, 
we have firstly to determine the probability that a randomly chosen square-free number is divisible by a prime factor $p_a$. We will show that such a probability is given by $1/(p_a+1)$ by using the inclusion-exclusion principle as follows. Let $F(t)$ be the number of square-free numbers which are less than $t$: this function asymptotically goes as 
\EQ
F(t) \,\simeq\, \frac{6}{\pi^2} \, t \,\,\,.
\EN 
Using $F(t)$, we can give the first estimate of the number of square-free numbers which are less than $t$ and multiples of the prime $p_a$. This number is approximatively equal to $
F\left(t/p_a\right) 
$. 
If we now multiply a square-free number $y$ (with $y \leq t/p_a$) by $p_a$, this yields (for sure) a multiple of $p_a$ which is $\leq t$. This multiplication usually gives rise to a number which is also square-free, for the only perfect square that could possibly divide the number $y p_a$, where $y \leq t/p_a$ and $y$ is square-free, is $p_a^2$. This implies that $F(t/p_a)$ overcounts the set of multiples of $p_a$ that are $\leq t$ and square-free. In order to correct this discrepancy, we must subtract approximately $F(t/p^2)$, which almost counts how many numbers $\leq t$ are divisible by $p_a^2$ but which are otherwise square-free. But this time we have subtracted too much, since we have also subtracted the numbers $\leq t$ which are divisible by $p_a^3$ but which are otherwise square-free. So, we need to add back approximately $F(t/p^3)$ and so on. In this way, we have to deal with the sum of the infinite series
\EQ
\frac{1}{p} - \frac{1}{p^2} + \frac{1}{p^3} - \frac{1}{p^4} + \cdots \,=\,\frac{1}{p+1}
\EN 
This yields the sought probability for a random chosen square-free number to be divisible by a prime $p_a$. To see how accurate this prediction is, we have generated the first $10^7$ square-free numbers and we have count how many of them were divisible by $3, 5, 7, \ldots$. The outputs of this analysis is in Table I 
and, as one can see, the agreement between ``theory'' and ``experiment'' is pretty remarkable.

\section{Probability of getting a ground state in the Hamiltonian ${\mathcal H}_R$}
Let's now compute the probability of getting one of the ground states of the Hamiltonian ${\mathcal H}_R$. 
As discussed in the text, this consists in estimating the probability of pairwise coprimality of ${\mathcal N}$ randomly selected square-free numbers. An important input of this computation is the probability that a square-free number is divisible by a prime factor $p_a$ which, as shown above, is given by $1/(p_a +1)$. With this information, we can follow the argument given in Schroeder's book [7]: the probability that none of ${\mathcal N}$ square-free integers has the prime factor $p_a$ is 
\EQ
\left(1 - \frac{1}{p_a +1}\right)^{\mathcal N} \,\,\,,
\EN 
while the probability that exactly one has $p_a$ as a factor is 
\EQ
\frac{{\mathcal N}}{p_a +1} \, \left(1 - \frac{1}{p_a +1}\right)^{{\mathcal N} -1} \,\,\,.
\EN 
The sum of these two probabilities is the probability that at most one of the ${\mathcal N}$ square-free numbers has $p_a$ as a factor. Hence, taking the product over all primes, we get the probability that ${\mathcal N}$ square-free numbers are pairwise coprime, i.e. the probability to get one of the ground states of the Hamiltonian (6) 
\EQ
{\rm Prob}({\rm ground\,states})\,=\, 
\prod_{p_a} \left[\left(1 + \frac{{\mathcal N}-1}{p_a +1}\right)\,\left(1 - \frac{1}{p_a +1}\right)^{{\mathcal N}-1}  \right] \,\,\,. 
\label{probabilitygroundstatesss}
\EN
This probability rapidly decreases by increasing the number ${\mathcal N}$ of chains of the ladder, as shown in Table II. In the limit ${\mathcal N} \rightarrow \infty$, 
the dimension of the ground state manifold is infinitesimally small with respect to the dimension of the Hilbert space.

\begin{table}[t]
\begin{center}
\begin{tabular}{|l|c|}\hline
${\mathcal N} $ & {\rm Prob} \\\hline
$3 $ & $0.511335$ \\
$4 $ & $0.299667$\\
$5$ & $0.160472 $\\
$6 $ & $0.0799262 $\\
$7$ & $0.0374877$ \\
$8 $ & $0.0167083$ \\
$9$ & $ 0.0071255$ \\ 
$10$ & $0.00292332$ \\ 
$50$ & $ 1.55 \times 10^{-24}$\\
$100$ & $7.74 \times 10^{-57}$\\
\hline
\end{tabular}
\end{center}
\label{tableggg}
\caption{Probability to get a ground state of the Hamiltonian (6) for different number of lattice sites ${\mathcal N}$.} 
\end{table}

\section{The competition between the hopping and the magnetic field}

Let's consider the 1-d Hamiltonian 
\begin{eqnarray}
&& H_i (h,\nu)\,=\, \sum_{a=1}^{\Lambda} h(a) F_i(a) + \\
&& \hspace{2mm} -  \frac{1}{2} \,\sum_{a \neq b} J_{ab} \, \left(f^\dagger_i(a) f_i(b) + f^\dagger_i(b) f_i(a) \right)
\,\,\,. \nonumber
\label{PotentialFinale11}
\end{eqnarray}
We want to address the following issue: once the behaviour of $h(a)$ is assigned, is there a simple condition on the hopping term  $J_{ab}$ which ensures that, in the limit $\Lambda \rightarrow \infty$, the two terms in this Hamiltonian are truly competing? What we mean is the following: once we parameterize the two couping constants as 
\EQ
h(a) \,=\, h \, \hat h(a) 
\hspace{3mm}
,
\hspace{3mm}
J_{ab} \,=\, \nu \,\hat J_{ab} 
\,,
\EN
is there the possibility to fix appropriately the dependence of $\hat J_{ab}$ (accordingly to the behaviour of $\hat h(a)$) in such a way that in the limit $\Lambda \rightarrow \infty$ the ratio $h/\nu$ remains finite and can be varied at our will?

In the following we assume that the magnetic field $h(a)$ grows in a way proportional to the logarithm of the prime numbers 
\EQ
h(a) \,=\, h\, \log p_a\,\,\,,
\label{propermagneticfield}
\EN
This is precisely the behaviour which ensures that a state as 
\EQ
| n_i \rangle \,=\,
\left(\prod_{a=1}^k (f_i^\dagger(a))^{\alpha_a} \right) \, |{\rm vac} \rangle \ \hspace{3mm}, 
\hspace{3mm} 
\,\,\,\alpha_a=\{0,1\}.
\label{local}
\EN
has a magnetization 
\EQ
M_{n_i} \,=h\, \, \sum_{a=1}^k \alpha_a \, \log (p_a)_i \,=\,h\, \log n_i \,\,\,.
\label{energyENii}
\EN
Therefore we can assign uniquely to a given value of $M/h$ a configuration of square-free numbers and viceversa. 

 Assuming that $J_{ab}$ is a circulant matrix, the eigenvectors of the hopping term alone is particularly simple 
and can be expressed in terms of the $\Lambda$-roots of the unity $ \omega \,=\,e^{2 \pi i/\Lambda}$, with $\omega^{ k n} \,=\, e^{(2 \pi i/\Lambda) k\, n}$ 
In fact, we have 
\[
|\phi_k \rangle 
\,=\,
\frac{1}{{\sqrt{\Lambda}}}\,
\left(
\begin{array}{c}
\omega^{0 k} \\
\omega^{1 k} \\
\omega^{2 k} \\
\cdot\\
\cdot\\
\cdot\\
\omega^{(\Lambda -1) k}  
\end{array}
\right)
\,\,\,, 
\label{complex}
\]
with $k=0,1,\ldots, \Lambda -1$. In particular, the expression for the ground state, which corresponds to the non-degenerate value $e_0$, is particularly simple 
\[
\hspace{-3mm}
| \phi_{0} \rangle \,=\,\frac{1}{\sqrt{\Lambda}} \,\left(
\begin{array}{c} 
1 \\
1\\
1\\
\cdot\\
\cdot\\
\cdot\\
1
\end{array}
\right)
\,\,\,.
\]
Expressing it as $|\phi_0\rangle =\sum_a c_a |v_a\rangle$ we can compute the ground state energy of the hopping term alone, given by 
\[
E_0 \,=\,\frac{1}{\Lambda} \sum_{a=1}^{\Lambda} \sum_{b=1}^{\Lambda} J_{ab} \,\,\,,
\label{ggg}
\]
Let's now image that we switch on a small magnetic field and let's then compute the first order correction induced by the magnetic field on this ground state. This correction is given by 
\[
\delta e_0 \,=\,\frac{1}{\Lambda} \sum_{a=1}^{\Lambda} \tilde h(a) \,\,\,.
\]
If we now impose that $\delta_0 \simeq E_0$, we arrive to the condition 
\EQ 
\sum_{a,b} J_{a,b} \simeq \sum_{a} \tilde h(a) \,\,\,.
\label{condition}
\EN
Notice that if $h(a)$ is a constant, $J_{ab}$ has to be short-range, the usual next neighborhood case being given by $b = a \pm 1$. 
If, on the other hand, $h(a)$ is an increasing function (as the one given in eq.\,(\ref{propermagneticfield}),  assuming $J_{a,b} = J{|a-b|}$ the condition (\ref{condition})
can be also expressed as 
\be
J(x) \simeq \hat h'(x)\,\,\,. 
\ee
This is the condition which fixes how $J_{ab}$ should be fixed in order that the two interaction terms in the Hamiltonian (\ref{PotentialFinale11}) are on the same footing. 
Since $\hat h(x) \,=\,\log p_x \simeq \log x$, this leads to $\hat J(x) \,\simeq \,1/x$. 

\begin{figure}[t]
\centering
\includegraphics[width=0.30\textwidth]{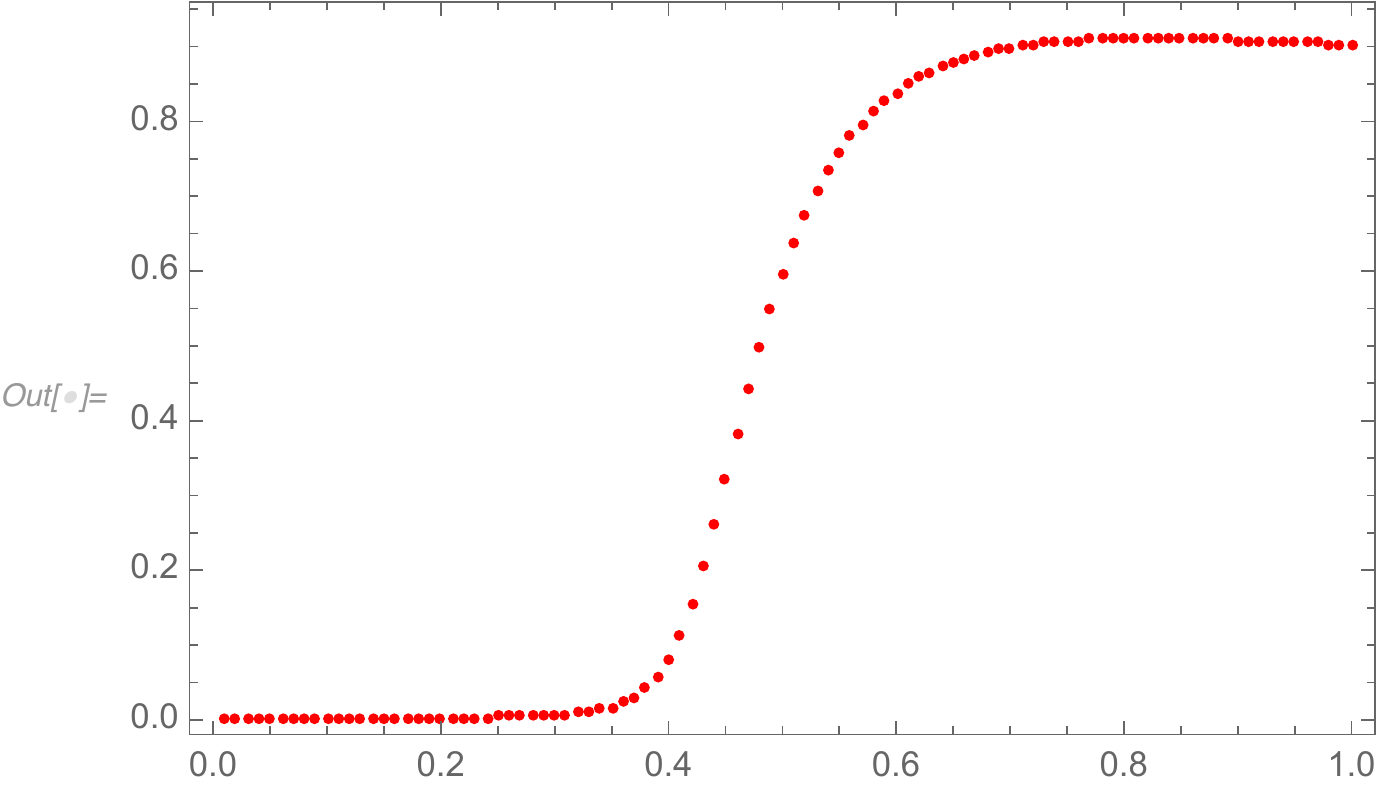}
\caption{IPR(x) vs $x$, with $x/(1-x) = \nu/h$, for the Hamiltonian ${\mathcal H}_i$, with $\hat h_a = \log p_a$ and $J_{ab}=1/|a-b|$
.}
\label{IPRniceplot}
\end{figure}

To check that this behaviour of $J_{ab}$ is the correct one to have a truly competitive situation between the two terms in the Hamiltonian we can make use of the Inverse Participation Ratio  of the ground states wave function in one particle sector 
$|\phi_0\rangle = \sum_{a=1}^\Lambda c_a |p_a \rangle$, defined as  ${\rm IPR}(x) = 1/(\Lambda \sum_a |c_a|^4)$, 
where $x/(1-x)\equiv \nu/h$. The IPR is indeed a good diagnosis of the transition between a localised phase (${\rm IPR} \simeq 0$, for $x\rightarrow 0$) and a delocalised phase (${\rm IPR} \simeq 1$, for $x \rightarrow 1$): the corresponding plot is in Figure \ref{IPRniceplot}. 

\begin{thebibliography}{XX}
\bibitem{footnote2} Note that for $\lambda =0$, the Hamiltonian was diagonalized by nested Bethe-ansatz in \cite{Sutherland}.
\end{thebibliography}
\end{document}